%% file: ms.tex
\documentclass[12pt]{article}

\usepackage{amsmath,amsthm,amssymb}
\usepackage{graphicx,psfrag,epsf}
\usepackage[caption=false]{subfig}
\usepackage{enumerate}
\usepackage{multirow,booktabs}
\usepackage{url} 
\usepackage{natbib}

\usepackage{multibib}
\newcites{refsupp}{Supplementary references}
\usepackage{longtable}

\usepackage{bm}
\usepackage[section]{placeins}
\makeatletter
\AtBeginDocument{%
	\expandafter\renewcommand\expandafter\subsection\expandafter{%
		\expandafter\@fb@secFB\subsection
	}%
}
\makeatother

\usepackage{etoolbox}
\makeatletter
\patchcmd{\@makecaption}
{\parbox}
{\advance\@tempdima-\fontdimen2} 
{}{}
\makeatother 

\usepackage{geometry}
\usepackage{titling}
\makeatletter
{\begin{titlepage}\def\@thanks{}}%
{\end{titlepage}}
\makeatother

\renewcommand{\vec}[1]{\mathbf{#1}}
\newcommand{\mat}[1]{\mathbf{#1}}
\newcommand\ddfrac[2]{\frac{\displaystyle #1}{\displaystyle #2}}

\newtheorem{theo}{Theorem}
\newtheorem{lem}[theo]{Lemma}
\newtheorem{prop}[theo]{Proposition}

\theoremstyle{remark}
\newtheorem{rmk}{Remark}

\providecommand{\keywords}[1]{\textbf{Keywords:} #1}

\title{Probabilistic $K$-mean with local alignment for clustering and motif discovery in functional data}

\author{Marzia A. Cremona\thanks{marzia.cremona@fsa.ulaval.ca} \\ 
		\small{Dept.~of Statistics, The Pennsylvania State University, University Park, USA} \\
		\small{Dept.~of Operations and Decision Systems, Université Laval, Canada} \\
		\small{CHU de Québec – Université Laval Research Center, Canada}
		\and
		Francesca Chiaromonte\thanks{fxc11@psu.edu} \\ 
		\small{Dept.~of Statistics, The Pennsylvania State University, University Park, USA} \\
		\small{Inst.~of Economics and EMbeDS, Sant’Anna School of Advanced Studies, Pisa, Italy}
	}

\begin{document}
	\maketitle
	
	\begin{abstract}
		We develop a new method to locally cluster curves and discover functional motifs, i.e.~typical ``shapes'' that may recur several times along and across the curves capturing important local characteristics. 
		In order to identify these shared curve portions, our method leverages ideas from functional data analysis (joint clustering and alignment of curves), bioinformatics (local alignment through the extension of high similarity seeds) and fuzzy clustering (curves belonging to more than one cluster, if they contain more than one typical ``shape''). 
		It can employ various dissimilarity measures and incorporate derivatives in the discovery process, thus exploiting complex facets of shapes.
		We demonstrate the performance of our method with an extensive simulation study, and show how it generalizes other clustering methods for functional data. 
		Finally, we provide real data applications to Berkeley growth data, Italian Covid-19 death curves and ``Omics'' data related to mutagenesis.
	\end{abstract}
	
	
	\section{Introduction}
	\label{sec:introduction}
	\input{introduction}
	
	\section{Probabilistic $K$-mean with local alignment}
	\label{sec:probKMA}
	\input{probKMA}
	
	\section{Cluster evaluation and functional motif discovery}
	\label{sec:FMD}
	\input{FMD}
	
	\section{Simulations}
	\label{sec:simulation}
	\input{simulation}

	\section{Real data applications}
	\label{sec:examples}
	\input{examples}

	\section{Discussion}
	\label{sec:discussion}
	\input{discussion}
	
	\section*{Supplementary material}
	The online Supplement contains proofs and additional details on post-processing, simulations and real-data analyses. 
	
	\section*{Acknowledgments}
	We thank Matthew Reimherr and Piercesare Secchi for discussions about functional data methodology; 
	Kateryna D. Makova and Di (Bruce) Chen for help with the mutagenesis application;
	Valeria Vitelli and Davide Floriello for their sparse functional clustering code. 
	
	\section*{Fundings}
	This work was partially funded by the Eberly College of Science, the Institute for Cyberscience and the Huck Institutes of the Life Sciences (Penn State University);
	NSF award DMS-1407639; 
	and Tobacco Settlement and CURE funds of the PA Department of Health. 
	M.A. Cremona acknowledge the support of the NSERC. 
	
	\bibliographystyle{Chicago}
	\small{\bibliography{references}}

\end{document}

%% file: introduction.tex
Given a set of curves, we consider the problem of discovering \emph{functional motifs} inside them, i.e.~typical ``shapes'' that may recur within each curve, and across several curves in the set. 
Some of these motifs may be present in most of the curves, but in different positions. Conversely, other motifs may characterize subgroups of curves -- and thus differentiate among them based on local shape similarities.
We provide a novel method for functional motif discovery that aligns curves locally to identify their shared portions, employing different definitions of (dis)similarity. 
Importantly, neither the motifs nor their number, lengths or radii need to be known in advance; in addition, lengths and radii are specific to each motif. 

During the last two decades, the analysis of curves has received increasing attention and interest in the statistical literature. Indeed, several functional data analysis (FDA) methods have been developed and applied in many fields \citep[see, e.g.,][]{ramsay2005,ferraty2006,horvath2012}. 
Several algorithms have been proposed to cluster aligned functional data \citep[reviewed in][]{jacques2014}. 
Since functional data are very often misaligned, algorithms have also been proposed to simultaneously cluster and align curves \citep{liu2009,sangalli2010,park2017}. 
All these methods consider the curves globally, over their entire domain of definition. However, in many applications, separation in groups may occur only on a portion of the domain; this type of clustering structure might be missed by methods that consider curves in their entirety. 
The multivariate counterpart of this ``domain selection'' problem 
is usually referred to as \emph{feature selection}, 
and has been widely studied  \citep[see, e.g.,][]{friedman2004,witten2010}.
In the FDA framework, \cite{fraiman2016} and \cite{floriello2017} proposed methods to cluster curves while performing feature (i.e.~domain) selection. 
More recently, \cite{vitelli2019} integrated curve alignment in the sparse clustering procedure. 

The problem of \emph{functional motif discovery} we tackle here is more general -- and to the best of our knowledge, it has never been studied in the statistical literature. In order to identify motifs, we define clusters locally on portions of the misaligned curves, and allow each cluster to contain multiple portions of the same curve (i.e.~multiple instances of the same functional motif). In addition, we allow each curve to belong to multiple clusters (i.e.~to comprise multiple functional motifs). 
This problem is the continuous version of sequence motif discovery, which is ubiquitous in 
bioinformatics and ``Omics'' sciences \citep[see, e.g.,][]{bailey2006} and consists of searching for highly similar patterns in a set of DNA or protein sequences. While these are discrete sequences of symbols (4 nucleotides, or 20 amino acids), we consider curves that can attain any real values -- and can in fact be multivariate (i.e.~take values in 
$\mathbb{R}^d$). 
A similar problem for time series has been addressed by the data mining community \citep{lin2002,mueen2009,yeh2016,yeh2018} 
defining a motif as a pattern repeated multiple times within a single time series. Available tools generally employ the Euclidean distance or the correlation between portions of the time series. 
They usually require as input the length and the number of motifs to be found, although \cite{linardi2018} recently introduced an algorithm that finds all motifs in a given range of lengths. 
Importantly, these tools require a user-specified minimum distance within which two portions of the time series are considered the same motif (i.e.~a ``motif radius''), and this distance is the same across motifs. 

We embed the problem of functional motif discovery in a full-blown FDA framework, which allows us to capture complex shape characteristics by incorporating derivatives in the discovery process. The FDA framework also allows us to rigorously define variability within each motif, and to naturally reduce noise in the curves through smoothing. 
Our novel method, \emph{probabilistic $K$-mean with local alignment} (probKMA), leverages ideas from FDA, bioinformatics and fuzzy clustering in order to identify $K$ shared curve portions, which represent $K$ candidate functional motifs in the set of curves under consideration. 
Indeed, similar to the $K$-mean with (global) alignment of \cite{sangalli2010}, we simultaneously perform clustering and alignment of curves. However, we employ local alignment in place of their global alignment. Also, similar to BLAST-type algorithms in bioinformatics \citep{blast}, we perform local alignments through the extension of high similarity ``seeds''. Finally, similar to fuzzy clustering in which points can belong to multiple clusters \citep{bezdek1981,bezdek1984}, our curves can be associated with zero, one, or more than one cluster (if they contain more than one typical ``shape''). 

The article is organized as follows. 
In Section~\ref{sec:probKMA} we present the theoretical setting of probKMA, formulate it as an optimization problem, derive necessary conditions for its solution 
and describe its algorithmic implementation. 
In Section~\ref{sec:FMD} we discuss evaluation of the clusters produced by the algorithm  
and identification of the motifs discovered. 
In Section~\ref{sec:simulation} we provide an extensive simulation study to evaluate probKMA and compare it to other approaches. 
Finally, we present real data applications in Section~\ref{sec:examples} and 
provide concluding remarks in Section~\ref{sec:discussion}.

%% file: probKMA.tex
We consider a set of $N$ ($d$-dimensional) curves \mbox{$\vec{x}_i:\ \mathbb{R} \longrightarrow \mathbb{R}^d$}, \mbox{$i=1,\dotsc,N$}. 
Our goal is to identify $K$ ($d$-dimensional) cluster centers \mbox{$\vec{v}_k:\ (0,c_k) \longrightarrow \mathbb{R}^d$}, \mbox{$k=1,\dotsc,K$} of unknown lengths $c_1, \dots, c_k \in [c_{min},c_{max}]$. 
These are ``patterns'' to which the curves are, locally, highly similar with respect to a distance $d(\cdot,\cdot)$. 
Since we are interested in local similarity (i.e.~similarity between portions of curves) and we define each cluster center only in the interval $(0,c_k)$, we allow each curve to be aligned to each cluster center as to minimize their distance. 
Alignment to each cluster center $\vec{v}_k$ is performed composing each curve $\vec{x}_i$ with a warping function \mbox{$h_{k,i} :\ \mathbb{R} \longrightarrow \mathbb{R}$} from a class $W$. 
Here we consider shifts \mbox{$W=\{h:\ t \longmapsto t+s; s \in \mathbb{R}\}$}, but our method can be generalized to other warping functions commonly employed in the FDA literature \citep[see, e.g.,][]{ramsay2005}. 

Because of the focus on local similarity, a curve can belong to more than one cluster; that is, different portions of a curve can be similar to portions of other curves. Hence, 
mimicking fuzzy clustering \citep[see, e.g.,][]{bezdek1981,bezdek1984}, we assign to each curve $\vec{x}_i$ a probability $p_{k,i}$ to be a member of each cluster $k$. In particular, we define a membership function $p_k:\ \{\vec{x}_1,\dotsc,\vec{x}_N\} \longrightarrow [0,1]$ for each $k=1,\dotsc,K$, with $p_k(\vec{x}_i)=p_{k,i}$, requiring that
$\sum_{k=1}^{K} p_{k,i}=1$ for all $i=1,\dotsc,N$, and that 
$\sum_{i=1}^N p_{k,i}>0$ for all $k=1,\dotsc,K$. 
Each membership probability $p_{k,i}$ corresponds to a particular shift $s_{k,i}$ of the curve $\vec{x}_i$; namely, the one that minimizes the distance between $\vec{x}_i$ and $\vec{v}_k$ given all constraints.
We summarize these shifts in a matrix \mbox{$\mat{S}=\left[s_{k,i}\right]$} and the membership probabilities in a matrix $\mat{P}=\left[p_{k,i}\right]$.

\subsection{Optimization problem and necessary conditions}
\label{subsec:optim}
	Consider the cluster center lengths $c_1, \dots, c_k$ as fixed (identification of $c_1, \dots, c_k \in [c_{min},c_{max}]$ is discussed in the next Subsection). 
	ProbKMA can be formulated as the following optimization problem: find $K$ cluster centers $\vec{v}_1,\dotsc,\vec{v}_K$, membership probabilities $\mat{P}$ and shifts $\mat{S}$ that minimize the generalized least-squares functional
	\begin{equation}
		\label{eq:Jm}
		J_m(\mat{P},\mat{S},\vec{v}_1,\dotsc,\vec{v}_K)=
			\sum_{i=1}^{N}{\sum_{k=1}^{K}{\left(p_{k,i}\right)^m d^2\left(\vec{\tilde{x}}_{i,s_{k,i}},\vec{v}_k\right)}}
	\end{equation}
	under the constraints $p_{k,i} \in [0,1]$, $\forall i,k$; $\sum_{k=1}^{K} p_{k,i}=1$, $\forall i$; and $\sum_{i=1}^N p_{k,i}>0$, $\forall k$. 
	Here \mbox{$m>1$} is a fixed weighting parameter controlling the degree of ``fuzziness'', and \mbox{$\vec{\tilde{x}}_{i,s_{k,i}}=\vec{x} \circ h_{k,i}$} are the shifted curves. 
	Necessary conditions for $(\hat{\mat{P}},\hat{\mat{S}},\hat{\vec{v}}_1,\dotsc,\hat{\vec{v}}_K)$ to be a (local) minimizer of (\ref{eq:Jm}) are that each of $\hat{\mat{P}}$, $\hat{\mat{S}}$ and $\hat{\vec{v}}_1,\dotsc,\hat{\vec{v}}_K$ minimizes (\ref{eq:Jm}) fixing all the other variables. 
	We prove two key results (see Supplementary Material).
	The first provides an explicit solution
	for the optimal membership probabilities $\hat{\mat{P}}$ given shifts and centers. 
	Importantly, this result holds for any distance $d(\cdot,\cdot)$ and does not rely on any regularity assumption on the curves or the cluster centers. 
	\begin{prop}
		\label{prop:memberships}
		Fix a shift matrix $\hat{\mat{S}}$ and $K$ cluster centers $\hat{\vec{v}}_1,\dotsc,\hat{\vec{v}}_K$. 
		Consider the set \mbox{$R=\left\{i \in \{1,\dotsc,N\}\, |\, d(\vec{\tilde{x}}_{i,\hat{s}_{k,i}},\hat{\vec{v}}_k)>0 \mbox{ for all } k \right\}$} and suppose that $|R| \geq K$. \\
		Then $\hat{\mat{P}}=\left[\hat{p}_{k,i}\right]$ is a global minimizer of the functional 
		\begin{equation}
			\label{eq:Jm_memberships}
			J_m(\cdot,\hat{\mat{S}},\hat{\vec{v}}_1,\dotsc,\hat{\vec{v}}_K)\ :\ [0,1]^{K,N} \longrightarrow \mathbb{R},
		\end{equation}
		under the constraints $\sum_{k=1}^{K} p_{k,i}=1$, $\forall i$ and $\sum_{i=1}^N \hat{p}_{k,i}>0$, $\forall k$ if and only if 
		\begin{equation}
			\label{eq:membershipsR}
			\hat{p}_{k,i}=\left[\sum_{l=1}^{K}{\left( 
				\frac{d^2\left(\vec{\tilde{x}}_{i,\hat{s}_{k,i}},\hat{\vec{v}}_k\right)}
							{d^2\left(\vec{\tilde{x}}_{i,\hat{s}_{l,i}},\hat{\vec{v}}_l\right)} 
			\right)^{\frac{1}{m-1}}}\right]^{-1} \qquad k=1,\dotsc,K
		\end{equation}
		for all $i \in R$ and 
		\begin{equation}
			\label{eq:membershipsnotR}
			\hat{p}_{k,i}=
			\begin{cases}
				0, & k \,:\, d(\vec{\tilde{x}}_{i,\hat{s}_{k,i}},\hat{\vec{v}}_k)>0 \\
				\in [0,1], & k \,:\, d(\vec{\tilde{x}}_{i,\hat{s}_{k,i}},\hat{\vec{v}}_k)=0
			\end{cases}
		\end{equation}
		with $\sum_{k=1}^K \hat{p}_{k,i}=1$, for all $i \notin R$. 
	\end{prop}
	If the $i$-th curve has positive distance from all cluster centers, (\ref{eq:membershipsR}) states that its probability of belonging to cluster $k$ is inversely proportional to its distance from the $k$-th cluster center. 
	Equation (\ref{eq:membershipsnotR}) tackles the extreme case of a curve with distance $0$ from one or more cluster centers; in this case the functional minimum is attained by setting the membership probabilities to $0$ for all clusters from which the curve has positive distance. Note that if a curve has distance $0$ to more than one cluster (\ref{eq:membershipsnotR}) defines an uncountable set of minimizers. However, the non-uniqueness is trivial. 
	
	The second result concerns updates of the cluster centers and depends on the distance employed. 
	This provides an explicit solution for optimal centers given shifts and probabilities 
	for any distance $d_{\alpha}(\cdot,\cdot)$ defined as 
	\begin{equation}
		\label{eq:dalpha}
		d_\alpha^2 \left( \vec{x},\vec{v} \right)=
		\sum_{j=1}^{d}{\frac{w_j}{d}}
		\Big[
			\frac{1-\alpha}{c} \int_0^c{\left( x^{(j)}(t)-v^{(j)}(t) \right)^2 \mathrm{d}t}
			+ \frac{\alpha}{c} \int_0^c{\left( x'^{(j)}(t)-v'^{(j)}(t) \right)^2 \mathrm{d}t}
		\Big],
	\end{equation}
	where $w_j>0$ is the weight of the $j$-th component of a $d$-dimensional curve, indicated by $^{(j)}$, $'$ indicates the weak derivative, and $\alpha \in [0,1]$ is a parameter that defines the relative weight of the curve's levels and derivatives. 
	Selecting $\alpha=0$ we obtain an $L^2$-like distance $d_0(\cdot,\cdot)$ that focuses exclusively on the levels. 
	In this case 
	we require $\vec{x}_i \in L^2(\mathbb{R},\mathbb{R}^d)$ and \mbox{$\vec{v}_k \in L^2((0,c_k),\mathbb{R}^d)$}.
	The choice of $\alpha=1$ leads to an $L^2$-like pseudo-distance $d_1(\cdot,\cdot)$ that uses only weak derivative information, hence focusing on curve variations (their slopes or trends). 
	Finally, $\alpha \in (0,1)$ properly defines a Sobolev-like distance $d_\alpha(\cdot,\cdot)$ that allows one to highlight more complex features of curve shapes, taking into account both levels and variations. 
	When $\alpha>0$, 
	we require $\vec{x}_i \in H^1(\mathbb{R},\mathbb{R}^d)$ and $\vec{v}_k \in H^1((0,c_k),\mathbb{R}^d)$.
	\begin{prop}
		\label{prop:motifs}
		Fix a membership matrix $\hat{\mat{P}}$ and a shift matrix $\hat{\mat{S}}$. 
		Consider the distance $d_\alpha(\cdot,\cdot)$ with $\alpha \in [0,1]$. 
		For $\alpha=0$, assume that \mbox{$\vec{x}_i \in L^2(\mathbb{R},\mathbb{R}^d)$} and \mbox{$v_k \in L^2((0,c_k),\mathbb{R}^d)$} for $k=1,\dotsc,K$. 
		For $\alpha>0$, assume that $\vec{x}_i \in H^1(\mathbb{R},\mathbb{R}^d)$ and $v_k \in H^1((0,c_k),\mathbb{R}^d)$ for $k=1,\dotsc,K$. 
		Then $(\hat{\vec{v}}_1,\dotsc,\hat{\vec{v}}_K)$ is the (unique) global minimizer of the functional 
		\begin{equation}
			\label{eq:Jm_motifs}
			J_m(\hat{\mat{P}},\hat{\mat{S}},\cdot)\ :\ V_1 \times \cdots \times V_K \longrightarrow \mathbb{R}
		\end{equation}
		if and only if 
		\begin{equation}
			\label{eq:motifs}
			\hat{\vec{v}}_k=
				\ddfrac{\sum_{i=1}^N(\hat{p}_{k,i})^m \tilde{\vec{x}}_{i,\hat{s}_{k,i}}}
							 {\sum_{i=1}^N(\hat{p}_{k,i})^m}
				\qquad \text{a.e. in } (0,c_k),\, \forall k.
		\end{equation}
		For $\alpha=1$, $\hat{\vec{v}}_k$ is defined by (\ref{eq:motifs}) up to an additive constant.
	\end{prop}
	Equation (\ref{eq:motifs}) defines the $k$-th cluster center has a weighted average of the shifted curves in the interval $(0,c_k)$. 
	Weights are determined by membership probabilities: the contribution of a curve to the computation of the $k$-th cluster center is directlt proportional to its probability of belonging to cluster $k$.

\subsection{Algorithm}
\label{subsec:algorithm}
	Propositions \ref{prop:memberships} and \ref{prop:motifs} suggest to numerically minimize (\ref{eq:Jm}) through an iterative procedure that alternates: \ref{algo:centers}.~\emph{identification of cluster centers} with equation (\ref{eq:motifs}), \ref{algo:align}.~\emph{curve alignment} (warping function selection), and \ref{algo:prob}.~\emph{computation of membership probabilities} using equations (\ref{eq:membershipsR})-(\ref{eq:membershipsnotR}). 
	We propose the following algorithm for probKMA. 
	\begin{description}
		\item[Initialization] Fix the number of clusters $K$ and the cluster center lengths $c_1,\dotsc,c_K$. 
		Consider an initial membership matrix $\mat{P}^{(0)}$ such that $\sum_{k=1}^{K} p_{k,i}^{(0)}=1$, $\forall i$ and $\sum_{i=1}^{N} p_{k,i}^{(0)}>0$, $\forall k$ (non-degenerate clusters), and an initial shift matrix $\mat{S}^{(0)}$; 
		\item[Iteration] For $it=1,2,\dotsc$, iterate the following three steps until convergence:
		\begin{enumerate}[i.]
			\item \vspace{-0.2cm} \emph{Identification of cluster centers.} 
			    For each $k$, compute the $k$-th cluster center $\vec{v}_k^{(it)}$ with equation (\ref{eq:motifs}), 
			    using the shift $s_{k,i}^{(it-1)}$ and membership probabilities $p_{k,i}^{(it-1)}$; 
			    \label{algo:centers}
			\item \emph{Curve alignment.} 
			    For each $i$ and $k$, align the curve $\vec{x}_i$ to the new cluster center $\vec{v}_k^{(it)}$, selecting the shift $s_{k,i}^{(it)}$ that minimizes their distance $d(\tilde{\vec{x}}_{i,s},\vec{v}_k^{(it)})$; \label{algo:align}
			\item \emph{Computation of membership probabilities.} 
			    Compute the membership matrix $\mat{P}^{(it)}$ with equations (\ref{eq:membershipsR})-(\ref{eq:membershipsnotR}), 
			    using $\vec{v}_k^{(it)}$ and 
			    the shifts $s_{k,i}^{(it)}$. \label{algo:prob}
		\end{enumerate}
		\item[Stopping criterion] At each iteration, the convergence of the probabilistic clustering is evaluated by means of the Bhattacharyya distance between the membership matrices $\mat{P}^{(it)}$ and $\mat{P}^{(it-1)}$.
		In particular, for each cluster $k$ we compute
		\begin{equation}
		    BC_k=-\log \left(\sum_{i=1}^N \sqrt{p_{k,i}^{(it)}p_{k,i}^{(it-1)}} \right). 
		    \label{eq:bhattacjaryya_k}
		\end{equation}
		The Bhattacharyya distance is computed as the maximum, mean, or order $q$ quantile of (\ref{eq:bhattacjaryya_k}) for all clusters $k$.
		Steps \ref{algo:centers}-\ref{algo:prob} are repeated until the global Bhattacharyya distance reaches a given tolerance. 
	\end{description}
	
	\begin{rmk}
		We observe that steps \ref{algo:centers} and \ref{algo:prob} are analogous to the steps of a fuzzy $K$-mean algorithm \citep{bezdek1984}, or to the steps of an EM algorithm for mixture models \citep{dempster1977}. 
		Also notably, steps \ref{algo:centers} and \ref{algo:align} correspond to the functional $K$-mean with (global) alignment algorithm \citep{sangalli2010}. 
	\end{rmk}
	
	Every iteration 
	can be written in functional form as 
	\begin{equation*}
		\left(\mat{P}^{(it)},\mat{S}^{(it)},\vec{v}_1^{(it)},\dotsc,\vec{v}_K^{(it)}\right)
		\in T_m\left(\mat{P}^{(it-1)},\mat{S}^{(it-1)},\vec{v}_1^{(it-1)},\dotsc,\vec{v}_K^{(it-1)}\right)
	\end{equation*}
	where $T_m\ :\ Y \longrightarrow Y$ is the point-to-set map defined by \ref{algo:centers}-\ref{algo:prob}, and $Y$ the subset of \mbox{$[0,1]^{K,N} \times \mathbb{R}^{K,N} \times V_1 \times \dotsc \times V_K$} that satisfies $\sum_{k=1}^{K} p_{k,i}=1$, $\forall i$ and \mbox{$\sum_{i=1}^N p_{k,i}>0$}, $\forall k$. 
	For each initialization, the algorithm generates a sequence of iterations 
	\begin{equation}
		\label{eq:iter}
		\left\{T_m^{(it)}\left(\mat{P}^{(0)},\mat{S}^{(0)},\vec{v}_1^{(0)},\dotsc,\vec{v}_K^{(0)}\right)\right\}_{it=1,2,\dotsc}.
	\end{equation}
	Below we show that the functional (\ref{eq:Jm}) is continuous and descends along (\ref{eq:iter}). 
	This is an important result, which mimics the one in \cite{hathaway1987} for fuzzy $K$-mean (its proof is provided in the Supplementary Material). 
	\begin{lem}
		\label{lem:Jm_cont}
		The functional $J_m: Y \longrightarrow \mathbb{R}$ is continuous. 
	\end{lem}
	\begin{theo}
		\label{theo:descent}
		Consider $\vec{y}^{(it-1)} \in Y$. Then for every $\vec{y}^{(it)} \in T_m(\vec{y}^{(it-1)})$ we have 
		\begin{equation}
			\label{eq:descent}
			J_m\left(\vec{y}^{(it)}\right) \leq J_m\left(\vec{y}^{(it-1)}\right), 
		\end{equation}
		i.e.~$J_m$ is a descent functional for $T_m$. 
		Moreover, $J_m$ descends strictly along the iterations 
		if $\vec{y}^{(it-1)} \notin \Omega$, where $\Omega \subseteq Y$ is the solution set of $\hat{\vec{y}}=(\hat{\mat{P}},\hat{\mat{S}},\hat{\vec{v}}_1,\dotsc,\hat{\vec{v}}_K) \in Y$ such that 
		\begin{eqnarray}
			J_m\left(\hat{\mat{P}},\hat{\mat{S}},\hat{\vec{v}}_1,\dotsc,\hat{\vec{v}}_K\right) \leq J_m\left(\mat{P},\hat{\mat{S}},\hat{\vec{v}}_1,\dotsc,\hat{\vec{v}}_K\right) & \forall \, \mat{P} \in [0,1]^{K,N} \label{eq:solP} \\[-5pt]
			& \sum_{k=1}^{K} p_{k,i}=1 \nonumber \\
			& \sum_{i=1}^N p_{k,i}>0; \nonumber
        \end{eqnarray}
        \begin{eqnarray}
			J_m\left(\hat{\mat{P}},\hat{\mat{S}},\hat{\vec{v}}_1,\dotsc,\hat{\vec{v}}_K\right) \leq J_m\left(\hat{\mat{P}},\mat{S},\hat{\vec{v}}_1,\dotsc,\hat{\vec{v}}_K\right) & \forall \, \mat{S} \in \mathbb{R}^{K,N}; \label{eq:solS}
        \end{eqnarray}
        \begin{eqnarray}
			J_m\left(\hat{\mat{P}},\hat{\mat{S}},\hat{\vec{v}}_1,\dotsc,\hat{\vec{v}}_K\right) < J_m\left(\hat{\mat{P}},\hat{\mat{S}},\vec{v}_1,\dotsc,\vec{v}_K\right) & \forall \, \vec{v}_k \in V_k \label{eq:solV} \\
			& \vec{v}_k \neq \hat{\vec{v}}_k \nonumber. 
		\end{eqnarray}
	\end{theo}
	\begin{rmk}
		Although the previous result does not guarantee that every sequence of iterations (\ref{eq:iter}) converges to a minimizer of the functional $J_m$, it is a necessary condition for convergence, and a desirable property for the algorithm.
	\end{rmk}
	
	\paragraph{Cluster center lengths identification.}
		In the previous theoretical results and algorithm, the lengths $c_1, \dots, c_K \in [c_{min},c_{max}]$ of the cluster centers 
		remain fixed along iterations. 
		However, we seek to identify local similarities even when the lengths of the ``matching'' curve portions are not known \emph{a priori}. 
		This problem has been already tackled by local sequence alignment methods in bioinformatics, whose goal is to find similar stretches of unknown lengths within a collection of nucleotides or amino acids sequences. 
		In this context, one of the most widely used algorithms is BLAST \citep{blast}.
		BLAST starts by finding short stretches shared by the sequences, and uses them as \emph{seeds}. 
		It then 
		extends the seeds on both sides in order to construct larger local alignments, stopping when the similarity score drops below a given threshold. 
		Borrowing this logic, we add a \emph{center elongation} step to our algorithm. 
		This step is performed only when the algorithm is reaching convergence, 
		to guarantee that we do not extend low-quality cluster centers. 
		We attempt elongation 
		on both the left and the right of each center, generating the elongated center with equation (\ref{eq:motifs}) applied to the correspondingly elongated set of curves. 
		For elongation to be acceptable, we require that the corresponding objective function 
		$ J_{m,k}(\mat{P},\mat{S},\vec{v}_1,\dotsc,\vec{v}_K)=
				\sum_{i=1}^{N}{\left(p_{k,i}\right)^m d^2\left(\vec{\tilde{x}}_{i,s_{k,i}},\vec{v}_k\right)}
		$
		decreases or that it increases less than a given threshold $\Delta_{J_{m,k}}$. 
		

%% file: FMD.tex

To evaluate a clustering produced by probKMA, we develop a \emph{generalized silhouette index}, similar to the one used in classic clustering \citep{rousseeuw1987}. 
Our index, which is defined for \emph{portions} of curves, measures how well each portion fits in the cluster it was assigned to. 
First, from every curve $\vec{x}_i$, we extract the portions that correspond to the clusters (after dicotomizing the membership probabilities $\mat{P}$ into $0$ and $1$, see Supplementary Material). 
Next, we compute the average distance $d_j(k)$ of each extracted portion $j=1,\dotsc,J$ 
from cluster $k$ as the mean of the distances between $j$ itself and all the portions attributed to cluster $k$. 
We define the \emph{intra-cluster} distance $a_j$ as the average distance of portion $j$ from the cluster $k_j$ it belongs to, i.e.~$a_j=d_j(k_j)$, and the \emph{inter-cluster} distance $b_j$ as the minimum of the average distances of portion $j$ from all the other clusters, i.e.~$b_j=\min_{k \neq k_j}d_j(k)$. 
The \emph{generalized silhouette index} for portion $j$ is a number in $[-1,1]$ defined as 
\begin{equation}
	s_j=\frac{b_j-a_j}{\max(b_j,a_j)}.
	\label{eq:silhouette}
\end{equation}
Large values of $s_j$ indicate that $j$ is appropriately assigned to its cluster, while low values indicate bad assignments. In particular, negative values signify that portion $j$ is closer to a cluster different from the one it was assigned to. 
For each cluster $k$, we then compute the \emph{cluster average silhouette index} $S_k$ as the average silhouette index across all the portions assigned to $k$. This measures the compactness of the cluster and hence its quality.
Finally, the \emph{overall average silhouette index} $S$, i.e.~the average of all $S_k$'s, measures the overall quality of the clustering.
Similar to classic clustering, silhouettes for portions, clusters and overall clustering are visualized in a \emph{silhouette plot} that facilitates their interpretation. 


Likewise other $K$-mean algorithms, probKMA finds a local minimum of the functional $J_m$ and its output depends heavily on initialization. 
If the goal is to \emph{cluster the curves} in $K$ groups \emph{based on local similarity}, we repeat the algorithm using different initializations (and possibly different initial lengths) and we select the solution with the lowest value of $J_m$. 
If the goal is \emph{functional motif discovery}, we run the algorithm multiple times with different initializations, cluster number and motif initial lengths, and form the set of candidate motifs taking the union of the resulting solutions.
We ``clean up'' this set of candidate motifs using generalized silhouette indices and motif number of occurrences. 
We then \emph{merge} candidate motifs that are very similar to each other, as they may in fact correspond to the same ``true motif'' identified by multiple runs of probKMA. 
Finally, given the resulting set of \emph{discovered motifs}, we utilize a \emph{motif search} algorithm to locate all their instances in the input curves. 
For each motif, we map all portions of the curves with distance lower than a given radius $R$ from the motif. Full details are provided in the Supplementary Material. 


%% file: simulation.tex

Generating curves comprising functional motifs in a non-trivial task, since we require motifs to be smoothly embedded in curves while allowing them to occur with noise. 
In order to do this, 
we take advantage of the flexibility provided by B-splines. 
We consider a B-spline basis $\{\Phi_l\}_{l=1}^L$ of order $n$, with equally spaced knots $t_1,\dots,t_{L-n+2}$, and define each curve as 
\begin{equation}
	x(t)=\sum_{l=1}^{L} c_l \Phi_l(t),
\end{equation}
where $c_l \in \mathbb{R}$, $l=1,\ldots,L$ are coefficients to be chosen. 
The order $n$ controls smoothness and complexity of $x$ ($x$ is a curve of class $C^{n-2}$ and a piece-wise polynomial of degree $n-1$). Higher orders provide more degrees of freedom, allowing one to generate curves with more complex shapes, 
and smoother at the knots. 
Each basis function $\Phi_l$ has compact support; specifically, it is $0$ outside an interval of length $nT$, where $T$ is the distance between two subsequent knots. As a consequence, we can define a functional motif of length $T$ fixing the values of $n$ coefficients $c_{\text{m},i},\dots,c_{\text{m},i+n-1}$ and repeating them multiple times within the same curve or across different curves. 
Longer motifs of length $2T,3T,\dots$ -- that may result in more complex shapes -- can be created in a similar way, fixing the values of $n+1,n+2,\dots$ subsequent coefficients. 
Since a single curve can embed more than one functional motif, as well as more than one occurrence of the same motif, we require motifs to be separated by at least one sub-interval $(t_i,t_{i+1})$ as not to be artificially merged (i.e.~we require at least $n$ background coefficients between them). 
Motif occurrences that are ``the same", both in shape and level, are generated adding Gaussian noise to the corresponding coefficients: $\tilde{c}_{\text{m},j}=c_{\text{m},j}+\epsilon_j$, \mbox{$\epsilon_j \stackrel{iid}{\sim} N(0,\sigma^2)$}. 
Motif occurrences that are ``the same" in shape but have different levels are obtained adding a constant $d_{\text{m}}$ to all the coefficients that define a single occurrence, choosing different constants for different occurrences: \mbox{$\tilde{c}_{\text{m},j}=c_{\text{m},j}+d_{\text{m}}+\epsilon_j$}, \mbox{$\epsilon_j \stackrel{iid}{\sim} N(0,\sigma^2)$}. 
Background coefficients $c_{\text{bg}} \in \left[a,b\right]$ are randomly generated as \mbox{$\left(c_{\text{bg}}-a\right)/b \stackrel{iid}{\sim} Beta(0.45,0.45)$}. The $Beta$ distribution allows us to create reasonably different backgrounds for both the curve $x(t)$ and its derivative $x'(t)$. 
With this flexible model we can generate functional data in several complex scenarios, varying curve and motif lengths, as well as variability, frequencies and positions of the occurrences of each motif.
	
\subsection{Functional motif discovery: varying curve length and noise in motifs}
\label{subsec:N20_K2-10-10}
	The aim of this simulation study is to demonstrate the performance of probKMA in discovering functional motifs embedded in a set of curves, and to examine the effects of increasing curve length and the noise level comprised in motif occurrences. We consider two different scenarios, with sets of curves embedding 
	(1) motifs that share both shapes and levels; or
	(2) motifs that share shapes but have different levels.
	 
	\begin{figure}[!t]
		\centering
		\subfloat[\label{subfig:ex_data_scenario1_motif1}]{\includegraphics[page=1,trim={0cm 0.15cm 0cm 0cm},clip,width=0.4\linewidth]{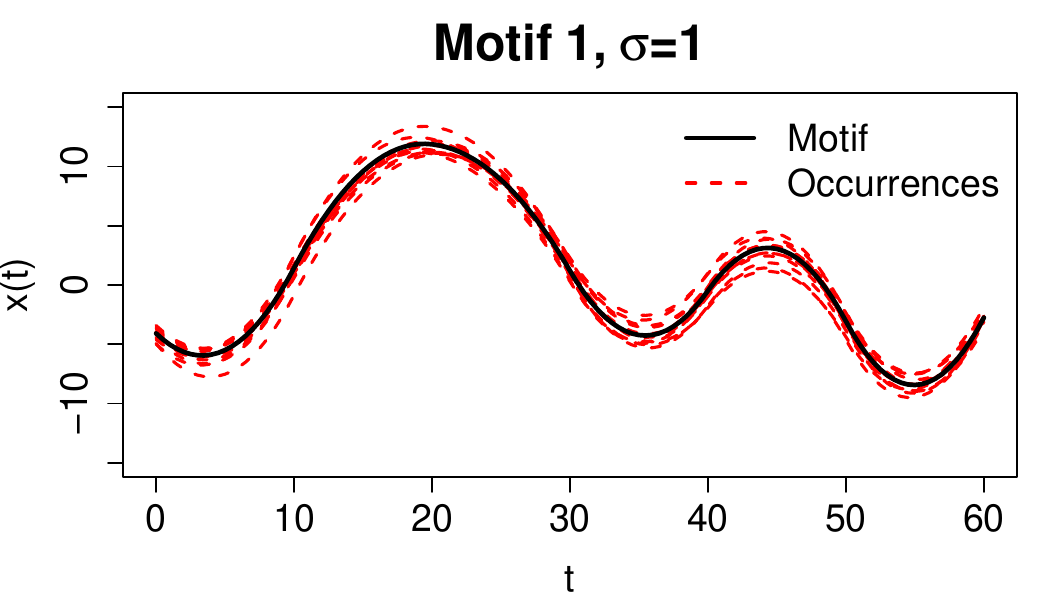}} 
		\subfloat[\label{subfig:ex_data_scenario1_motif2}]{\includegraphics[page=2,trim={0cm 0.15cm 0cm 0cm},clip,width=0.4\linewidth]{Figures/ex_motifs_with_noise1.pdf}}\\
		\subfloat[\label{subfig:ex_data_scenario1_curves_and_motifs}]{\includegraphics[trim={0cm 0.15cm 0cm 0cm},clip,width=0.8\linewidth]{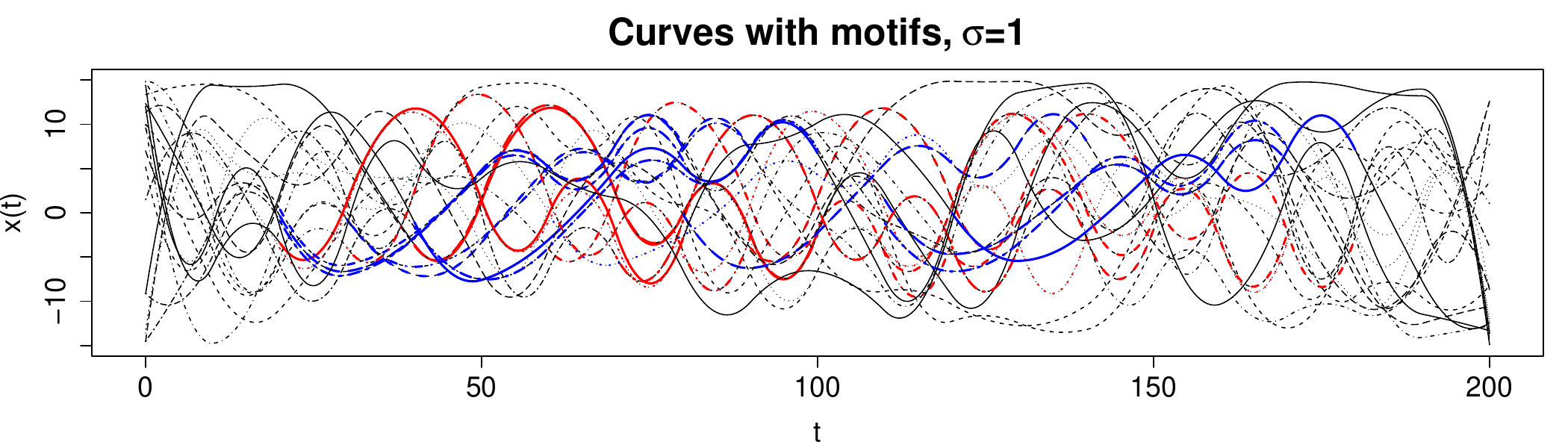}}
		\caption{
			Simulation scenario (1) with $l=200$ and $\sigma=1$. \protect\subref{subfig:ex_data_scenario1_motif1}, \protect\subref{subfig:ex_data_scenario1_motif2} Two functional motifs (black solid lines) and $12$ aligned occurrences of each (red and blue dashed lines); 
			\protect\subref{subfig:ex_data_scenario1_curves_and_motifs} $20$ curves embedding occurrences of the two motifs (red and blue portions, respectively).
		}
		\label{fig:data_scenario1}
	\end{figure}
	
	\begin{figure}[!hp]
		\centering
		\subfloat[\label{subfig:ex_results_scenario1_motif1}]{\includegraphics[page=1,trim={0cm 0.15cm 0cm 0cm},clip,width=0.6\linewidth]{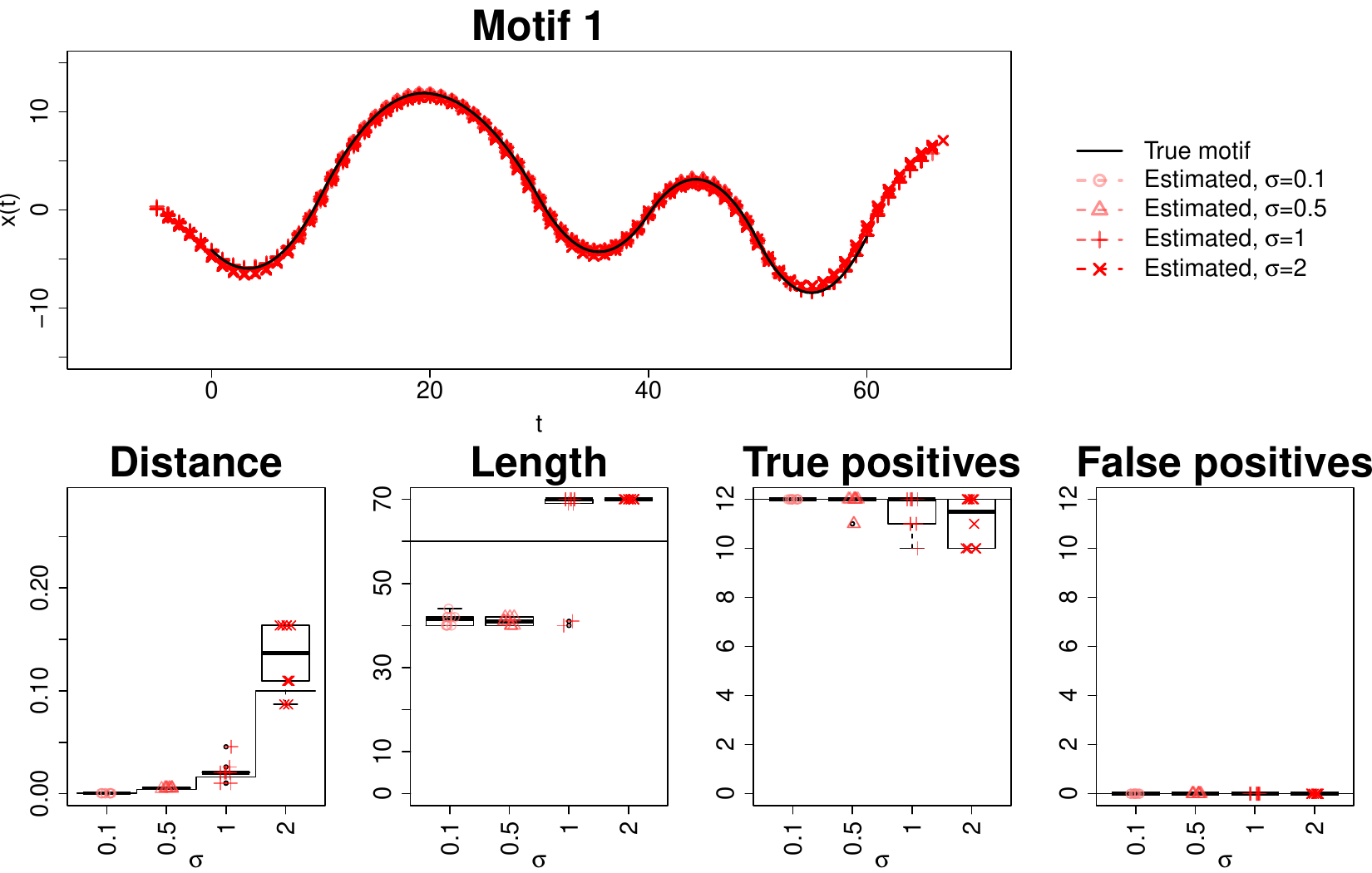}} \\
		\subfloat[\label{subfig:ex_results_scenario1_motif2}]{\includegraphics[page=2,trim={0cm 0.15cm 0cm 0cm},clip,width=0.6\linewidth]{Figures/ex_results_200.pdf}}
		\caption{
			Functional motif discovery results for simulation scenario (1) with $l=200$ and various levels of $\sigma$. 
			\protect\subref{subfig:ex_results_scenario1_motif1} Motif 1;
			\protect\subref{subfig:ex_results_scenario1_motif2} Motif 2. 
			The boxplots in the lower half of the panels are obtained from $10$ replications at each $\sigma$ value. They show the distance between true and estimated motifs (stepwise line: distance between the true motif and the average of all motif occurrences), the estimated length of motifs, the number of true positives and that of false positives. For all the considered noise levels, exactly 2 motifs are found.
		}
		\label{fig:results_scenario1}
	\end{figure}
	
	In scenario (1), we consider a set of $20$ curves embedding two functional motifs -- each with $12$ occurrences (see Fig.~\ref{fig:data_scenario1} and Supplementary Material). 
	In particular, $12$ curves contain only one occurrence of a motif ($6$ curves for each of the two motifs), $4$ curves contain two occurrences of a motif ($2$ curves for each of the two motifs), $2$ curves contain one occurrence of each of the two motifs, and $2$ curves contain no motif occurrences at all. 
	We generate data using a B-spline basis of order $3$, knots at distance $10$ and motifs of length $60$. Coefficients defining the two motifs are randomly generated from a $Beta(0.45,0.45)$ distribution rescaled to $\left[-15,15\right]$. 
	We consider four different curve lengths $l=200, 300, 400, 500$ and four levels of noise $\sigma=0.1, 0.5, 1, 2$ -- a total of $16$ simulated datasets. 
	In order to maximize the consistency among these datasets and thus highlight the effects of different $l$ and $\sigma$ values, we place motif occurrences within the leftmost sub-interval of length $200$ of each curve -- that is common to all datasets -- utilizing the same motif positions and background in all $16$ cases. Curves are sampled on a grid of points at distance $1$, so that each motif corresponds to $61$ points.
	For each combination of $l$ and $\sigma$, we run our probKMA-based functional motif discovery with Sobolev-like distance $d_{0.5}$. 
	We evaluate the number of motifs found, the distance between true and estimated motifs, the estimated lengths of motifs, and the number of true and false positives. 
	ProbKMA is run for $K=2, 3$, minimum motif lengths $c=40, 50, 60$ and $20$ random initializations for each $(K,c)$ pair (the maximum motif length is set to $70$; see Supplementary Material for other parameters). 
	The same initializations are employed for all $l$ and $\sigma$ combinations. 
	Simulation results for $l=200$ can be found in Fig.~\ref{fig:results_scenario1} and show very good performance for our method. 
	As expected, performance slightly declines when more noise is introduced in the motif instances: some occurrences can be missed, and/or false positives can be included. However, results remain satisfactory even when $\sigma=2$. 
	Results for other curve lengths are shown in the Supplementary Material. They suggest the same behavior as the noise level increases and, interestingly, they appear rather robust across lengths. 
	The only effect of increasing the ratio between ``background'' curve portions and curve portions occupied by motifs is a slight increase in false positives -- which occurs exclusively when also the noise level is high.

	In scenario (2) we consider exactly the same curves and motifs as in scenario (1), but allow motif occurrences to have different levels (see Supplementary Material). In particular, a random value $d_m \sim U(-10,10)$ is added to the coefficients defining an individual motif occurrence. 
	For each combination of $l$ and $\sigma$, we run our probKMA-based discovery with the $L^2$-like pseudo-distance $d_1$ (this lets us focus on curve variation). Parameters are the same as in scenario (1), and Figures detailing the results are provided in the Supplementary Material. 
	We find again that our method has good performance -- and that this performance is affected (as expected) by the noise level, but not much by the length of the curves. In some cases, especially when the curves are very long, we actually ``discover'' motifs that were not embedded in the simulated data.
	Note that, strictly speaking, these motifs are not all together ``false''. As one elongates the background portions of the curves, it is possible to generate by chance a few patterns that recur often enough to be identified by our algorithm. In our experiments, these additional motifs are noisier and have fewer occurrences than the two motifs originally embedded in the data.
	
	To validate the results described above, we repeat simulations in both scenarios $10$ times, considering $10$ different randomly generated pairs of motifs and re-generating the curves' background (see Supplementary Material). In all cases, our method shows good performance and similar behaviors as curve length and noise level change. 
	This is evidence that its effectiveness does not depend on on the specific shapes of the motifs embedded in the curves. 
	%
	Finally, we consider the $10$ simulations for the first scenario and examine how results change with the number of random initializations used to run probKMA for each $(K,c)$ pair. 
	To do this, we subsample the probKMA runs from the analysis already conducted, which employed $20$ random initializations -- using only $5$, $10$ or $15$ initializations for each $(K,c)$ pair. We re-run the post-processing steps and compare results with those previously obtained with all $20$ initializations.
	Reassuringly, our method is pretty robust to the number of initializations employed, and retains its good performance even when probKMA is run only $5$ times for each choice of $(K,c)$ (see Supplementary Material).
	
\subsection{Comparison with time series motif discovery}
\label{subsec:comparison_time_series}
	Next, we compare our probKMA-based functional motif discovery to time series motif discovery. As noted in the Introduction, the problem of motif discovery in a time series is similar to that of functional motif discovery. However, some important differences exist. 
    First, the \emph{definition of motif} is different. While our functional motifs 
	recur across a set of curves, and possibly within individual curves in the set, time series motifs are defined as recurring within a single time series -- usually starting from pairs of highly similar subseries
	(see Supplementary Material). 
	Second, time series motif similarity is based on Euclidean distance or correlation, while functional motif similarity is more general, based on a distance between functions that can incorporate derivatives. 
	Third, and perhaps most importantly, available time series motif discovery tools are not statistical in nature and they do not estimate the level of noise associated with each motif. The user needs to provide one, and usually \emph{only} one, motif radius as input to the discovery procedure. On the contrary, probKMA-based discovery ``learns'' an appropriate radius for each motif from the data. 
	
	\begin{table}[t]
		\caption{\label{tab:comp_matrix_profile} Comparison of probKMA-based functional motif discovery and Matrix Profile on simulation scenario (1)
			(TP: true positives; FP: false positives). For probKMA, we report median results (and median absolute deviations) across $10$ repeated simulations.}
		\centering \small
		{\def\arraystretch{0.8}
			\begin{tabular}{llrcccccccccccc}
				\addlinespace[0.2cm]
				&  &  & probKMA & \multicolumn{11}{c}{Matrix Profile} \\
				\cmidrule(lr{3pt}){4-4} \cmidrule(l{3pt}r){5-15}
				\multicolumn{3}{r}{Radius} & --- & 1 & 10 & 20 & 30 & 40 & 50 & 70 & 90 & 110 & 130 & 150 \\[-0.1cm] 
				\cmidrule(lr){1-15}
				& \multirow{2}{*}{Motif 1} & TP & \textbf{12 (0)} & 2 & 6 & 8 & \textbf{12} & 12 & 12 & 12 & 12 & 12 & 12 & 12 \\
				$l=200$ &  & FP & \textbf{0 (0)} & 0 & 0 & 0 & \textbf{0} & 0 & 0 & 0 & 0 & 0 & 0 & 0 \\ 
				\cmidrule(lr{3pt}){4-4} \cmidrule(l{3pt}r){5-15}
				$\sigma=0.1$ & \multirow{2}{*}{Motif 2} & TP & \textbf{12 (0)} & 2 & 7 & 10 & \textbf{12} & 12 & 12 & 12 & 12 & 12 & 12 & 12 \\ 
				&  & FP & \textbf{0 (0)} & 0 & 0 & 0 & \textbf{0} & 0 & 0 & 0 & 0 & 0 & 0 & 0 \\ 
				\cmidrule(lr){1-15}
				& \multirow{2}{*}{Motif 1} & TP & \textbf{11 (0.7)} & 0 & 0 & 0 & 0 & 0 & 0 & 0 & 1 & 2 & 2 & 2 \\
				$l=500$ &  & FP & \textbf{2 (1.5)} & 2 & 5 & 8 & 8 & 9 & 10 & 13 & 17 & 19 & 24 & 27 \\
				\cmidrule(lr{3pt}){4-4} \cmidrule(l{3pt}r){5-15}
				$\sigma=2$ & \multirow{2}{*}{Motif 2} & TP & \textbf{12 (0)} & 2 & 2 & 2 & 12 & 12 & 12 & 12 & 12 & 10 & 12 & 12 \\
				&  & FP & \textbf{1 (1.5)} & 0 & 1 & 2 & 8 & 16 & 23 & 34 & 51 & 71 & 82 & 88 \\
				\cmidrule(lr){1-15}
			\end{tabular}
		}
	\end{table}
    Among existing tools for time series motif discovery, we consider the recent \emph{Matrix Profile} \citep[][see Supplementary Material]{yeh2016,yeh2018}. 
    This tool discovers the top \emph{ motif pairs} in a time series and, for each of these pairs, provides all the ``neighboring'' subseries,
    i.e.~all the subseries 
    with distance less than $R$ from the motif pair. The same radius $R$ is used for all the top motif pairs. 
    We consider again the two simulation scenarios introduced in Subsection~\ref{subsec:N20_K2-10-10}, 
    focusing on two specifications: the simple case of short curves and low noise level ($l=200$ and $\sigma=0.1$), and the complex case of long curves and high noise level ($l=500$ and $\sigma=2$). 
    For probKMA-based discovery, we use the same parameters as in Subsection~\ref{subsec:N20_K2-10-10}. 
    We run Matrix Profile on one ``time series'' obtained by concatenating the 20 curves one after the other, and using different choices of radius (from $R=1$ to $R=150$). Since the tool requires also the motif length as input, we use the true motif length $c=60$ \cite[a newer implementation of Matrix Profile, introduced in][can find all motif pairs in a given range of lengths]{linardi2018}. 
	Table~\ref{tab:comp_matrix_profile} reports results for simulation scenario (1). Both probKMA-based motif discovery and Matrix Profile discover the two motifs in the simple case ($l=200$, $\sigma=0.1$). 
	However, when curves are longer and motifs noisier 
	(complex case, $l=500$, $\sigma=2$), Matrix Profile fails to find Motif 1, and includes many false positives in Motif 2. When the radius is large ($R>100$), it does correctly identify a small number of occurrences of Motif 1, but it also reports a very large number of false positives (for both motifs). 
	On the contrary, even in this complex case, probKMA-based motif discovery remains robust to noise level and curve length, and is able to identify both motifs with a very small number of false positives. 
	Similar observations can be made for simulation scenario (2) (see Supplementary Material).
	
\subsection{Comparison with non-sparse and sparse functional clustering}
\label{subsec:comparison_fda}
    Here we perform simulation experiments to compare probKMA, meant as a clustering method and separate from its motif discovery purpose, to other 
    functional clustering methods. 
    In particular, we include in the comparison the standard functional $K$-means \citep{tarpey2003} and $K$-mean with (global) alignment \citep[KMA][]{sangalli2010} -- which are both non-sparse methods -- and a recent sparse clustering technique \citep{floriello2017}. 
	
	\begin{figure}[!hp]
		\centering
		\subfloat[\label{subfig:ex_data_comp_only_motifs}]{\includegraphics[page=1,trim={0cm 0.15cm 0cm 0cm},clip,width=0.4\linewidth]{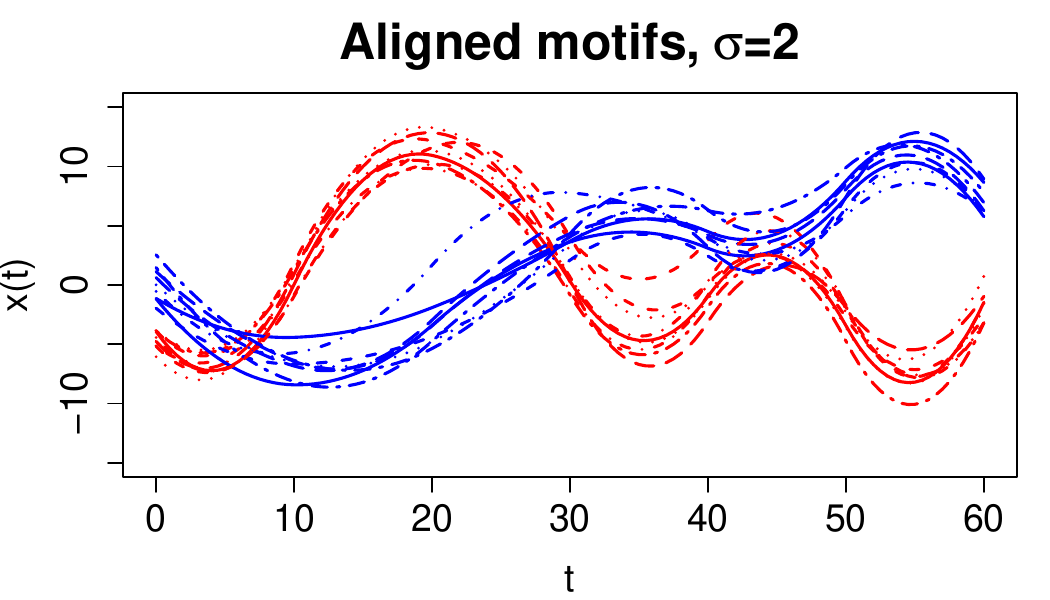}}
		\subfloat[\label{subfig:ex_data_comp_only_motifs_mis}]{\includegraphics[page=1,trim={0cm 0.15cm 0cm 0cm},clip,width=0.4\linewidth]{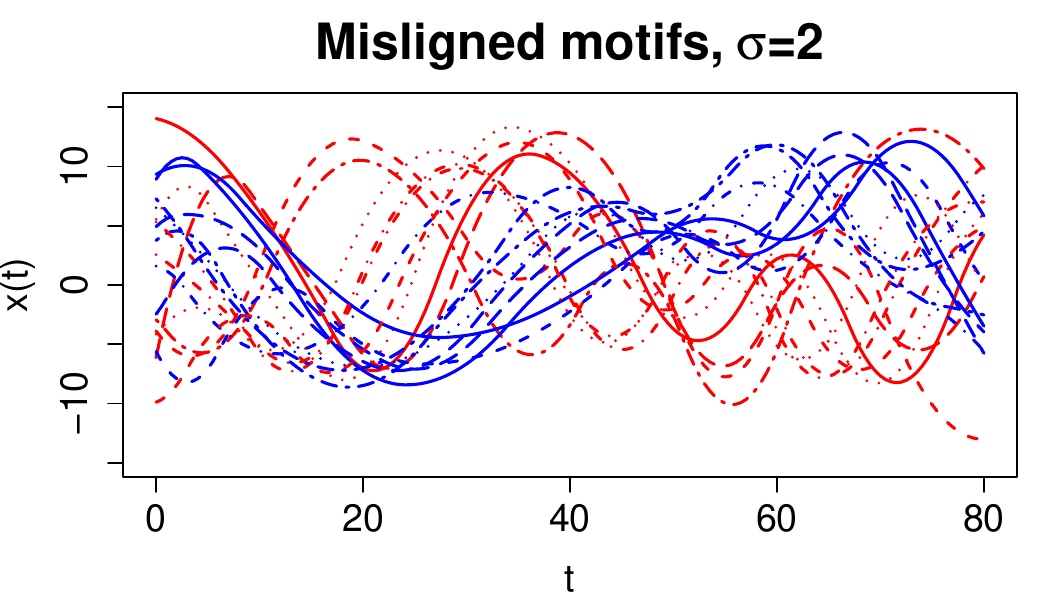}}\\
		\subfloat[\label{subfig:ex_data_comp_aligned_motifs}]{\includegraphics[trim={0cm 0.15cm 0cm 0cm},clip,width=0.8\linewidth]{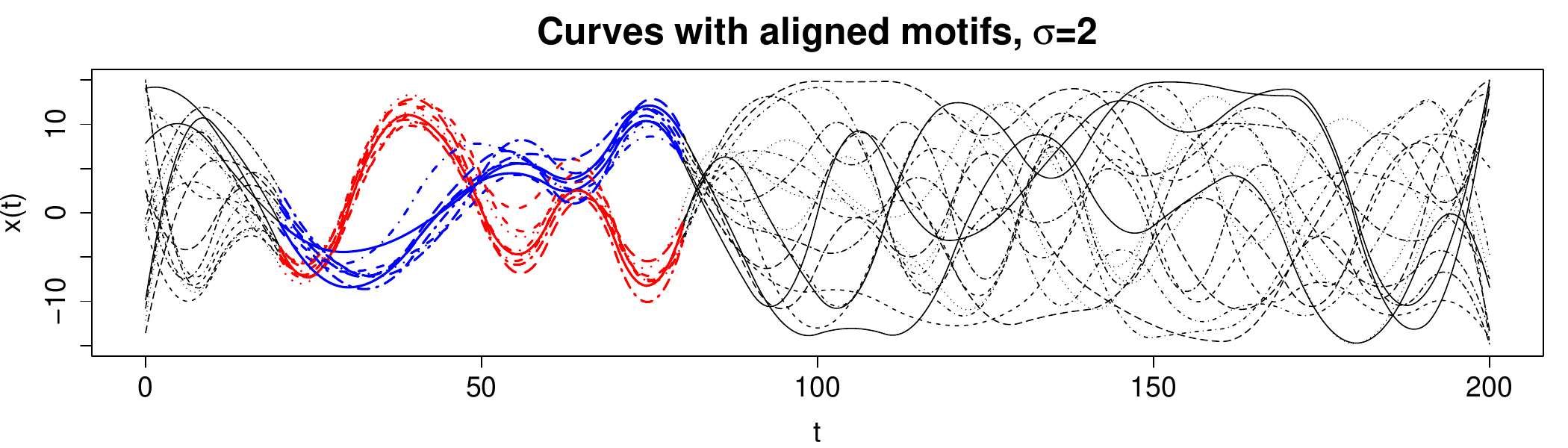}}\\
		\subfloat[\label{subfig:ex_data_comp_misaligned_motifs}]{\includegraphics[trim={0cm 0.15cm 0cm 0cm},clip,width=0.8\linewidth]{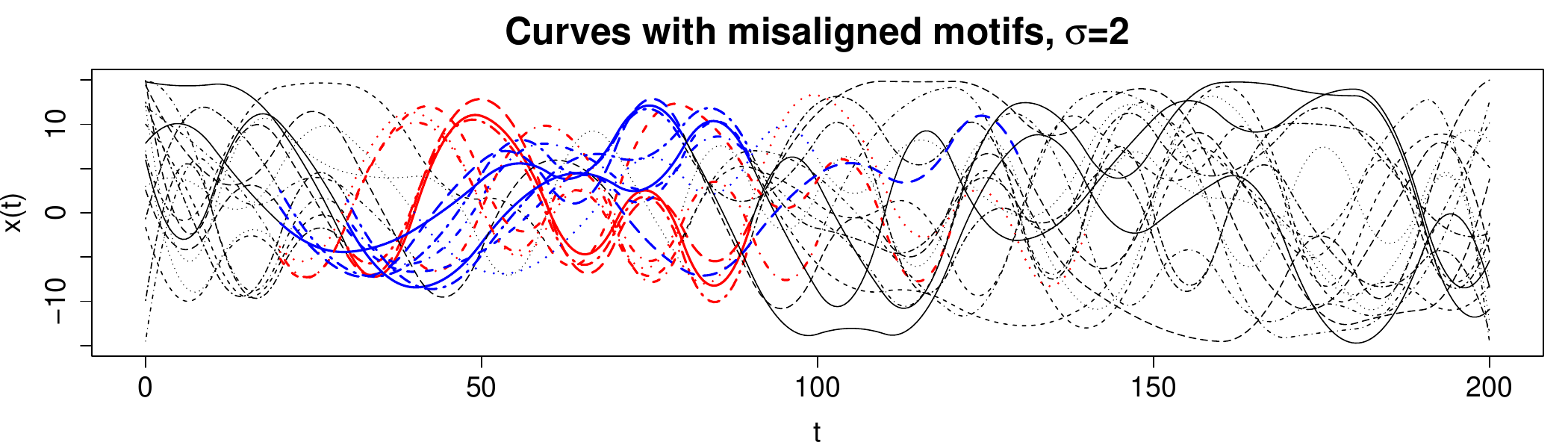}}
		\caption{
			Simulated data for the comparison of functional clustering methods, $\sigma=2$.
			\protect\subref{subfig:ex_data_comp_only_motifs} Aligned motifs;
			\protect\subref{subfig:ex_data_comp_only_motifs_mis} Misaligned motifs;
			\protect\subref{subfig:ex_data_comp_aligned_motifs} Curves with aligned motifs within them; 
			\protect\subref{subfig:ex_data_comp_misaligned_motifs} Curves with misaligned motifs within them.
			When the curves are broader than the motifs defining the two clusters, the motifs are shown as red and blue lines and the reminder of the curves as black lines.
		}
		\label{fig:data_comp}
	\end{figure}
	    \begin{table}
		\caption{\label{tab:comp_clustering} Comparison of probKMA with non-sparse and sparse functional clustering methods in four simulation scenarios. We report means (and standard deviations) of classification error rates across $10$ repetitions.}
		\centering \small
		{\def\arraystretch{0.8}
			\begin{tabular}{rccccc}
				\addlinespace[0.2cm]
				\multicolumn{2}{r}{Scenario} & $K$-means & KMA & sparse & probKMA \\ 
				\cmidrule(lr){1-6}
				\multirow{4}{*}{$\sigma=0.1$} & \subref{subfig:ex_data_comp_only_motifs} & \textbf{0 (0)} & \textbf{0 (0)} & \textbf{0 (0)} & \textbf{0 (0)} \\
				& \subref{subfig:ex_data_comp_only_motifs_mis} & 0.26 (0.13) & 0.12 (0.19) & 0.08 (0.18) & \textbf{0 (0)} \\
				& \subref{subfig:ex_data_comp_aligned_motifs} & 0.29 (0.22) & 0.44 (0.08) & 0.05 (0.17) & \textbf{0.04 (0.07)} \\
				& \subref{subfig:ex_data_comp_misaligned_motifs} & 0.49 (0.05) & 0.49 (0.06) & 0.52 (0) & \textbf{0.01 (0.04)} \\
				\cmidrule(lr){1-6}
				\multirow{4}{*}{$\sigma=2$} & \subref{subfig:ex_data_comp_only_motifs} & \textbf{0 (0)} & \textbf{0 (0)} & \textbf{0 (0)} & \textbf{0 (0)} \\
				& \subref{subfig:ex_data_comp_only_motifs_mis} & 0.26 (0.18) & 0.16 (0.21) & 0.42 (0) & \textbf{0 (0)} \\
				& \subref{subfig:ex_data_comp_aligned_motifs} & 0.28 (0.23) & 0.38 (0.17) & 0.11 (0.22) & \textbf{0.04 (0.10)} \\
				& \subref{subfig:ex_data_comp_misaligned_motifs} & 0.44 (0.07) & 0.49 (0.05) & 0.53 (0.01) & \textbf{0.06 (0.08)} \\
				\cmidrule(lr){1-6}
			\end{tabular}
		}
	\end{table}
	
	We consider $2$ clusters and generate $9$ curves for each cluster, in the four different scenarios depicted in Fig.~\ref{fig:data_comp}: 
	\subref{subfig:ex_data_comp_only_motifs} curves in the two clusters are aligned, and they differ on the entire domain; 
	\subref{subfig:ex_data_comp_only_motifs_mis} curves in the two clusters are misaligned, and they differ on the entire domain; 
	\subref{subfig:ex_data_comp_aligned_motifs} curves in the two clusters differ on a portion of the domain, and this portion is aligned; 
	\subref{subfig:ex_data_comp_misaligned_motifs} curves in the two clusters differ on a portion of the domain, and this portion is misaligned. 
	These four scenarios can be seen as special cases of the more general functional motif discovery problem, in which each curve contains exactly one motif and 
	\subref{subfig:ex_data_comp_only_motifs} curves are themselves the entire aligned motifs; 
	\subref{subfig:ex_data_comp_only_motifs_mis} curves are themselves the entire misaligned motifs; 
	\subref{subfig:ex_data_comp_aligned_motifs} curves contain aligned motifs;
	\subref{subfig:ex_data_comp_misaligned_motifs} curves contain misaligned motifs.
	In each scenario, we run all four clustering methods with Euclidean distance and $K=2$. 
    For the sparse clustering method, we also take the sparsity parameter (i.e.~the minimum length of the unselected part of the domain) as known, setting it to the curve length minus the motif length. We set the motif length parameter in probKMA in the same way. 
    In KMA and probKMA, we consider only shift alignments. 
    We then evaluate clustering results by means of a \emph{classification error rate} \citep[1 minus the Rand index;][]{rand1971} that is equal to $0$ if every curve is correctly classified and (since $K=2$) is equal to $0.5$ if the classification is as good as random. Since probKMA produces a probabilistic clustering, we compute the classification error rate after assigning each curve to the cluster with highest membership probability. 
	Results (Table~\ref{tab:comp_clustering}) show that all four methods correctly classify all curves in scenario \subref{subfig:ex_data_comp_only_motifs}. 
    $K$-means only works in this scenario, while KMA performs well in scenarios \subref{subfig:ex_data_comp_only_motifs} and \subref{subfig:ex_data_comp_only_motifs_mis}, and sparse clustering performs well in scenarios \subref{subfig:ex_data_comp_only_motifs} and \subref{subfig:ex_data_comp_aligned_motifs}. Interestingly, when the noise level is small, sparse clustering also achieves a good performance in scenario \subref{subfig:ex_data_comp_only_motifs_mis}. 
    ProbKMA performs very well in all four scenarios. 

%% file: examples.tex
\subsection{Berkeley growth curves}
\label{subsec:growth}
The Berkeley Growth Study dataset (provided within the R package \verb fda ) consists of the heights of 39 boys and 54 girls recorded from age 1 to 18. We estimate the curves using monotone B-spline smoothing \citep[with order 6, knots at observed ages, roughness penalty on third derivative and smoothing parameter $\lambda=1/\sqrt{10}$,][]{ramsay2009}. 

First, we perform a global probabilistic K-means running probKMA with $K=2$ and $L^2$-like pseudo-distance $d_1$ between the entire curves (no alignment permitted). Assigning each curve to the cluster with highest membership probability, we obtain clusters that differ on the main pubertal growth spurt timing and roughly corresponds to boys and girls groups (boys grow later), with 11 misclassified children (2 boys and 9 girls, classification error rate 0.21). Notably, $K$-means (with $K=2$ and Euclidean distance between curve derivatives) produces exactly the same clusters. However, our probabilistic approach permit us to visualize curves whose membership is uncertain and to check the probabilistic memberships of misclassified children (see Supplementary Material). 

\begin{figure}[!ht]
	\centering
	\subfloat[\label{subfig:growth_results_c51_misaligned}]{\includegraphics[page=2,trim={0cm 0.15cm 0cm 0cm},clip,width=0.49\linewidth]{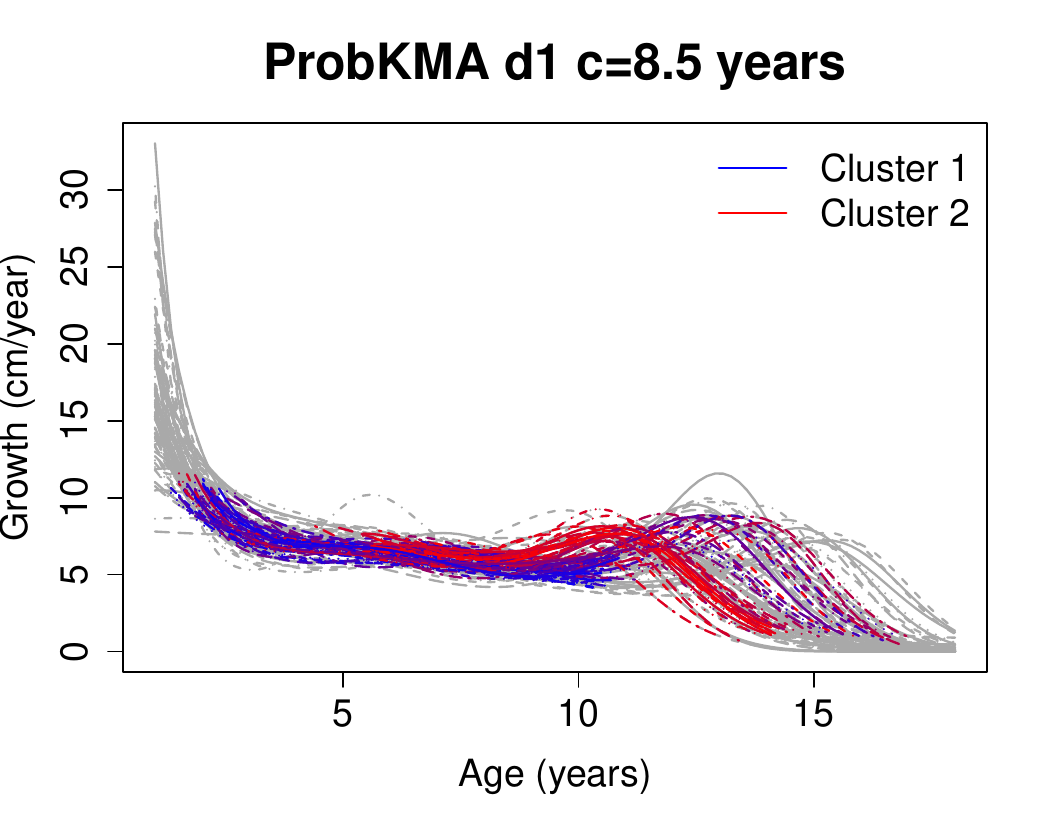}}
	\subfloat[\label{subfig:growth_results_c51_aligned}]{\includegraphics[page=8,trim={0cm 0.15cm 0cm 0cm},clip,width=0.49\linewidth]{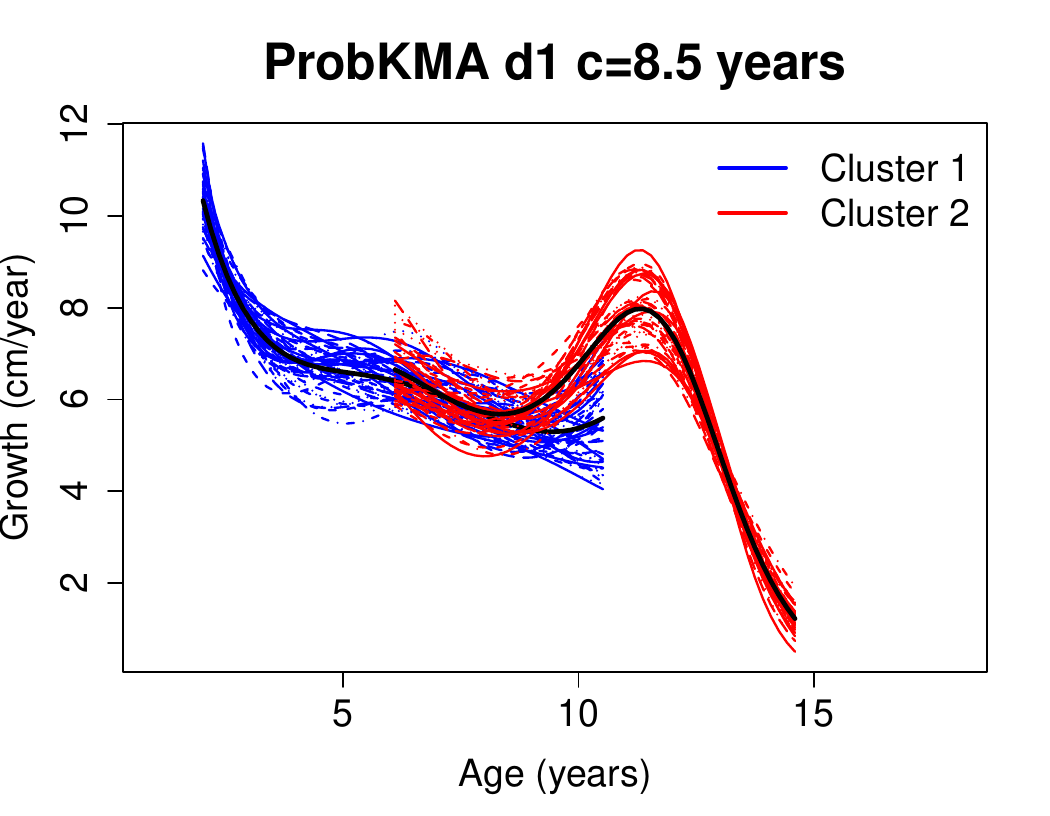}}
	\caption{
		Results of local clustering with probKMA ($K=2$). 
		\protect\subref{subfig:growth_results_c51_misaligned} Growth velocities with clustered portions of curves color-coded based on the probabilistic membership to Cluster 1 (going from red when it is 0 to blue when it is 1);
		\protect\subref{subfig:growth_results_c51_aligned} Cluster centers (black) with aligned clustered portions of curves.
	}
	\label{fig:growth_results_c51}
\end{figure}

Second, we cluster curves locally using probKMA with $K=2$ and $L^2$-like pseudo-distance $d_1$ between portions of curves of length 8.5 years, hence allowing for a maximum shift of 8.5 years. Dicotomizing the membership probabilities by setting them to 1 when the distance between the portion of curve and the cluster center is lower than the median distance, we obtain two clusters of 50 and 43 portions, respectively (Fig.~\ref{fig:growth_results_c51} and Supplementary Material). Interestingly, 18 curves belong to both clusters (i.e. they comprise both motifs in different parts of their domains), while 18 curves do not belong to any cluster. 
Cluster 2 captures a particular shape for the pubertal growth spurt, while Cluster 1 captures the decrease in growth velocity that is typical in children between 2 and 3 years of age.

\subsection{Italian Covid-19 excess mortality curves}
\label{subsec:covid}
\begin{figure}[!hp]
	\centering
	\subfloat[\label{subfig:covid_data}]{\includegraphics[trim={0cm 0.2cm 0cm 0cm},clip,width=0.8\textwidth]{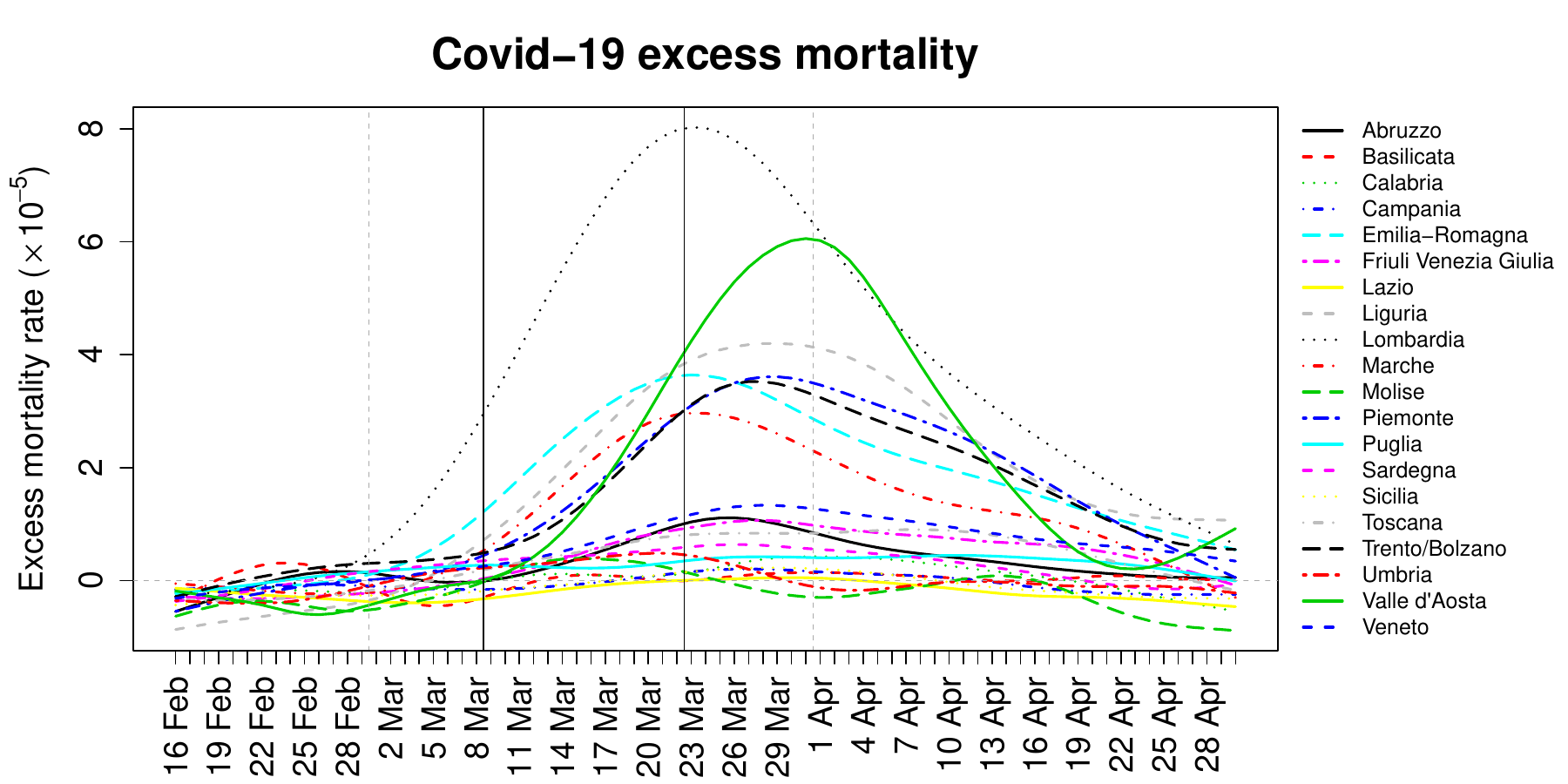}} \\
	\subfloat[\label{subfig:covid_results_aligned}]{\includegraphics[page=3,trim={0cm 0.2cm 0cm 0cm},clip,width=0.49\textwidth]{Figures/regions_istat_surplus_results.pdf}}
	\subfloat[\label{subfig:covid_results_alignment}]{\includegraphics[trim={0cm 0.2cm 0cm 0cm},clip,width=0.49\textwidth]{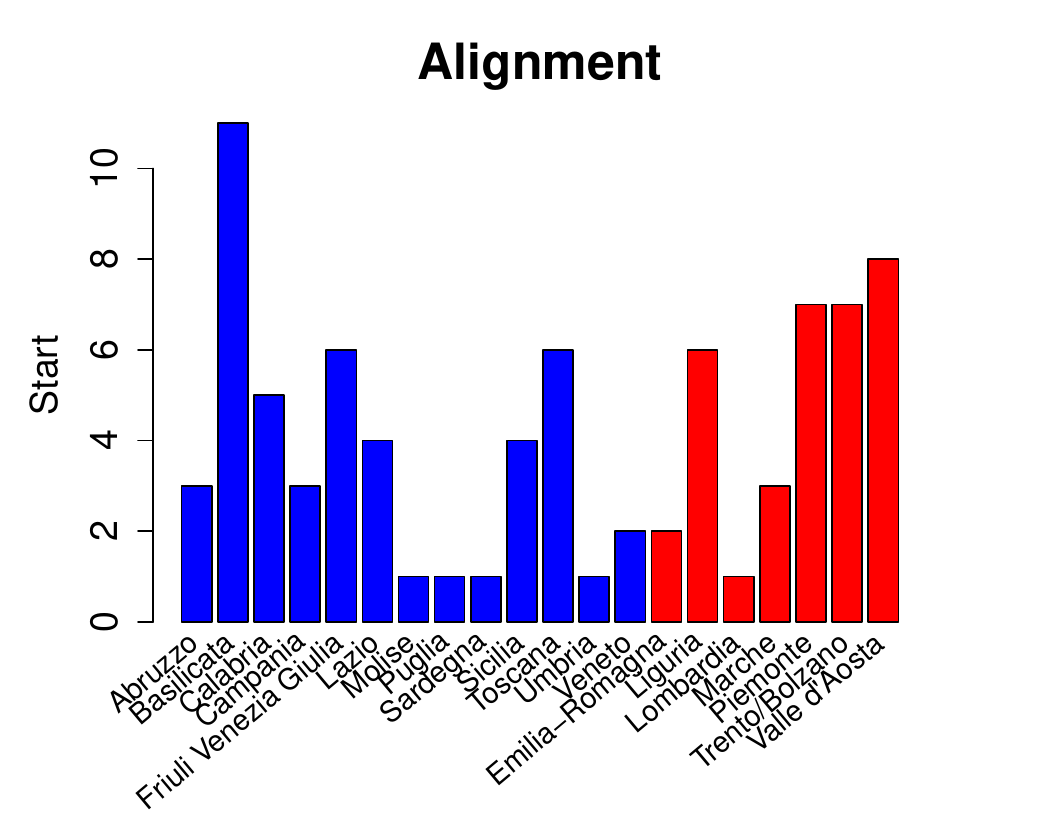}}
	\caption{
		Italian Covid-19 excess mortality rate curves and probKMA results. 
		\protect\subref{subfig:covid_data} Smoothed curves. Vertical black lines represent national lock down (March 9th) and closure of all non-essential economic activities (March 23rd);
		\protect\subref{subfig:covid_results_aligned} Cluster centers (black) with aligned clustered portions of curves;
		\protect\subref{subfig:covid_results_alignment} Alignment between portions of curves within each cluster (start day of each portion).
	}
	\label{fig:covid_results}
\end{figure}

Italy was the first European country to be hit by the Covid-19 pandemic, with the first confirmed cases around mid-February.
Italian regions were hit at different times and with different strength, and 
local authorities implemented different responses -- especially in the initial stages of the epidemic. Comparing its evolution across regions 
can therefore provide important insights on the role of underlying factors and different containment measures.
We estimate excess mortality due to Covid-19 in Italy using the mortality data (due to all causes) from the Italian Institute of Statistics (ISTAT). The dataset contains the daily number of deaths for 7.270 municipalities (covering about 93.5\% of the Italian population) from January 1st to April 30th, for the years 2015-2020. 
We aggregate data by region and we compute the excess mortality rate curves as the daily difference between 2020 deaths and average deaths in the period 2015-2019, divided by the population of the considered municipalities (see Supplementary Material). 
In order to focus on the Covid-19 period, we only consider data starting from February 16th. 
To reduce noise, we smooth the 20 curves using B-spline smoothing (cubic splines, knots at each day, roughness penalty on second derivative and smoothing parameter chosen by average Generalized Cross-Validation). Smoothed curves are shown in Fig.~\ref{fig:covid_results}\protect\subref{subfig:covid_data}. 

Next, we cluster the 20 Italian regions according to their excess mortality rate curves, to assess whether there are regions sharing similar epidemic patterns. 
We are interested in the entirety of the curves (possibly excluding the extremes of their domains), 
but we allow shifts in their alignment to take into consideration possible differences in 
the time when the (shared) patterns begun in each region. 
In particular, we employ probKMA as a clustering method, with $L^2$-like distance $d_0$ and cluster centers of length $c=65$ days (hence allowing for a maximum shift of 10 days). 
Fig.~\ref{fig:covid_results}\protect\subref{subfig:covid_results_aligned}-\protect\subref{subfig:covid_results_alignment} show probKMA results for $K=2$, when assigning each curve to the cluster with highest membership probability. 
Cluster 2 contains the regions (mainly located in the north of Italy) where 
Covid-19 hit the hardest. Lombardia is the region with earliest Covid-19 related deaths, followed by Emilia Romagna, Marche, Liguria, Piemonte and Trento/Bolzano, and last Valle d’Aosta (with a delay of 7 days). 
Cluster 1 contains the regions with milder epidemic patterns. 
Interestingly, Veneto is placed in Cluster 1 despite being the first region -- together with Lombardia -- to officially report Covid-19 cases. This suggests that Veneto successfully managed to ``flatten the curve'' with its early mass testing and contact tracing response \citep{mugnai2020}. 
In contrast, the pattern in Lombardia is so stark that it does not seem to 
fit properly even in Cluster 1 -- with a large distance from the cluster center (see Supplementary Material). Indeed, repeating the probKMA clustering with $K=3$, Lombardia is placed in a cluster of its own, while the other two clusters and the alignments within them do not change (see Supplementary Material).

\subsection{Mutagenesis data}
\label{subsec:mutagenesis}
To fully illustrate the proposed method in its motif discovery purpose, we apply it to a mutagenesis dataset adapted from \cite{kuruppumullage2013}.
Mutagenesis comprises all the processes by which mutations are generated in DNA, it is one of major evolutionary forces, and is central to causing many human diseases (e.g.~cancer). Understanding mutagenesis and how it is influenced by the genomic landscape is key to shedding light on genome dynamics \citep{makova2015}. 
\cite{kuruppumullage2013} estimated different types of neutral (i.e.~not affected by selection) mutation rates in non-overlapping windows 
along the human genome comparing it with primates, and employed Hidden Markov Models to define six divergence states and segment the genome accordingly. 
One of the states is of particular 
interest: it comprises ``hot regions'' with very high rates for substitutions, small insertions and deletions, which are associated with high GC content, early replication timing and open chromatin. 
Since these results were obtained at a rather large scale (1-Mb windows), investigating rates at finer resolution within the hot regions may reveal more specific trends and patterns of variation. 
Estimating high-resolution mutation rates in 1-kb windows within each hot region \citep[with the same pipeline as in][see Supplementary Material]{kuruppumullage2013} 
we generate a dataset of $43$ curves, varying in length from 1 Mb (corresponding to a grid of $1\,000$ points) to 22 Mb ($22\,000$ points). 
The curves are very noisy, and contain several missing or inaccurate values -- due to the fact that in many 1-kb windows 
the information needed to estimate rates is scarce. In particular, substitution rates can be reliably estimated only in $60\%$ of the 1-kb windows. 
After pre-processing with stochastic regression imputation and local smoothing, missing values are reduced to $17\%$ of the windows (see Supplementary Material). 
In this application we do not consider insertion and deletion rates, because their estimates are yet noisier and less accurate than for substitution rates. 

\begin{figure}[!b]
	\centering
	\subfloat[\label{subfig:substitution_results_motifs}]{\includegraphics[page=2,trim={0cm 0.15cm 0cm 0cm},clip,width=0.49\linewidth]{Figures/substitution_results_motifs.pdf}}
	\subfloat[\label{subfig:substitution_results_landscape}]{\includegraphics[page=1,trim={0cm 0.15cm 0cm 0cm},clip,width=0.49\linewidth]{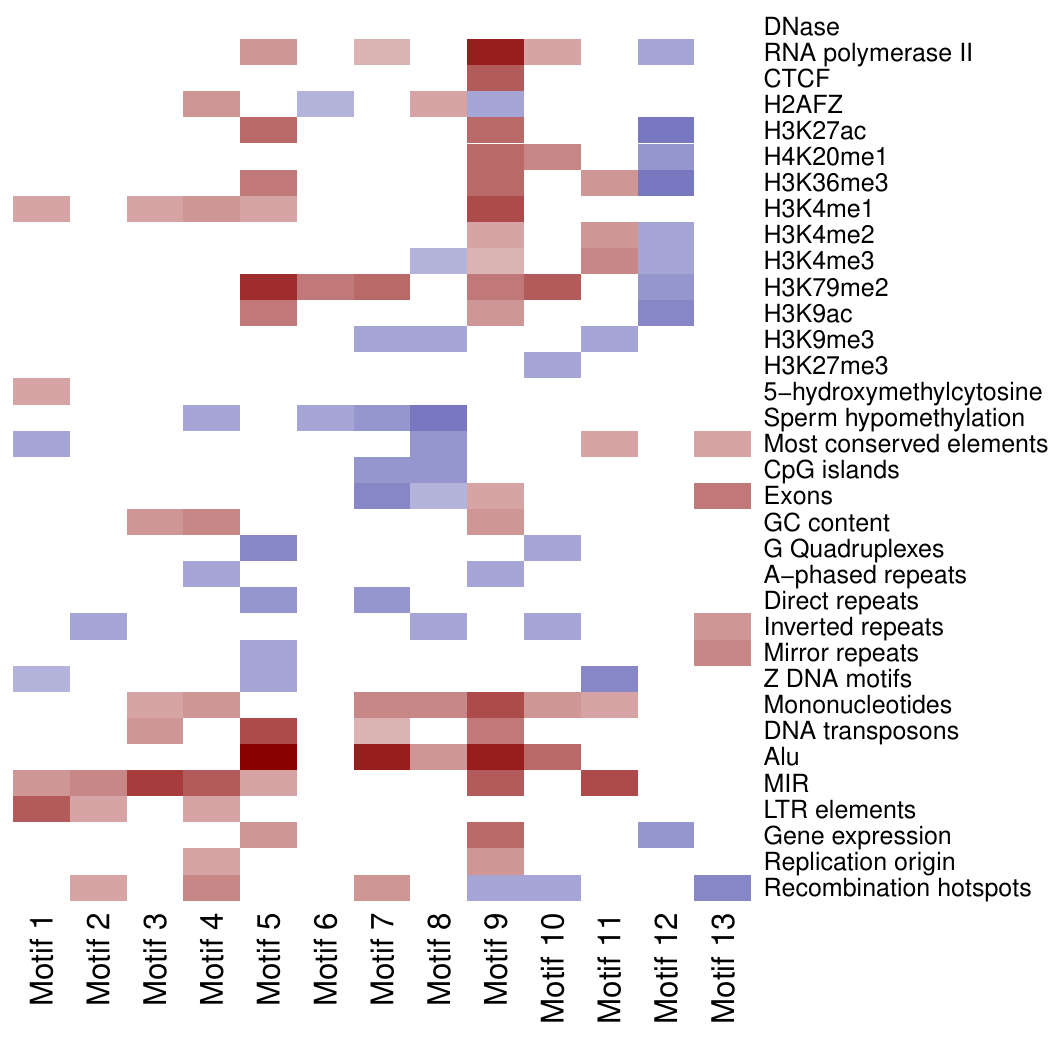}}
	\caption{
		ProbKMA-based functional motif discovery in substitution rate curves. 
		\protect\subref{subfig:substitution_results_motifs} Motifs found, plotted as percent changes with respect to the mean substitution rate across all hot regions;
		\protect\subref{subfig:substitution_results_landscape} Genomic landscape of the motifs, with color intensity proportional to the significance ($-\log_{10}(p)$) of a mean difference two-sided test contrasting motif occurrences and 
		hot regions at large -- red, blue and white represent positive, negative and non significant ($p>0.1$) differences, respectively.
	}
	\label{fig:substitution_results}
\end{figure}

\begin{table}[!ht]
	\caption{\label{tab:substitution_results} ProbKMA-based functional motif discovery in substitution rate curves. For each motif found, we report number of occurrences in the data and mean distance of the occurrences from the 
		motif.}
	\centering \small
	{\def\arraystretch{0.8}
		\begin{tabular}{rccccccccccccc}
			\addlinespace[0.2cm]
			Motif & 1 & 2 & 3 & 4 & 5 & 6 & 7 & 8 & 9 & 10 & 11 & 12 & 13 \\ 
			\cmidrule(lr){1-14}
			Number & 19 & 12 & 27 & 37 & 63 & 12 & 72 & 47 & 14 & 11 & 9 & 8 & 6 \\
			Mean dist & 1.9 & 1.9 & 3.5 & 5.1 & 6.5 & 3.0 & 8.7 & 7.4 & 5.2 & 3.0 & 5.5 & 18.5 & 17.0 \\
			\cmidrule(lr){1-14}
		\end{tabular}
	}
\end{table}

We employ our probKMA-based functional motif discovery on the $43$ curves using the Sobolev-like distance $\tilde{d}_{0.5}$ (the generalized version which can accommodate large gaps; see Supplementary Material). 
We look for motifs with minimum lengths $c=40,50,60,70$ (maximum length $c_{max}=150$), and we run probKMA for $K=2,3,4,5$ using $10$ random initialization for each $(K,c)$ pair. 
We employ our generalized silhouette index to evaluate each probKMA run and to filter the set of candidate motifs, and we select motif-specific radii based on probKMA results (see Supplementary Material). 
We identify $13$ functional motifs that differ substantially in length ($40$ to $104$ kb), levels and shapes (see Fig.~\ref{fig:substitution_results}\subref{subfig:substitution_results_motifs}). 
The motifs also differ substantially in frequency (i.e.~number of occurrences in the data) and level of variability (see Table \ref{tab:substitution_results}). This highlights the advantage of employing a motif discovery methodology able to learn motif-specific lengths, frequencies and variabilities based on the data. 

At least four of the motifs found are of biological interest:
Motif 12 corresponds to eight long sub-regions (about $100$ kb) with extremely high substitution rates -- an elevation of $10\%$ to $20\%$ relative to the mean level across all hot regions, which is already elevated in comparison to the genome at large. 
Motif 4 and Motif 8 also present very high substitution rates, and opposite patterns. 
In Motif 4, rates are about $10\%$ above the overall hot regions mean for the initial ${\sim}20$ kb's, and then decrease. 
In Motif 8 rates increase and then stabilize at about $10\%$ above the mean for ${\sim}20$ kb's. 
The two motifs have similar variability and are both very frequent ($37$ and $47$ occurrences, respectively).
Finally, Motif 13 corresponds to six long sub-regions with a substitution rate $20{\text -}30\%$ below the mean. These portions of the hot regions are in fact not hot; substitutions rates are similar to those of the rest of the genome.  
To investigate the genomic landscape of the motifs found, we consider a set of $35$ genomic features -- measured in each of the 1-kb windows constituting the hot regions. These features represent biological contexts that have an interplay with mutagenesis, such as DNA conformation, DNA sequence, replication, recombination, chromatin openness and modifications (see Supplementary Material). 
We then compare, independently for each genomic feature and each 
motif, the mean of the measurements in motif occurrences with the mean across all hot regions. In particular, we perform a simulation-based two-sided test for mean difference, where the empirical null distribution is obtained from $1000$ datasets generated by randomly relocating motif occurrences within the set of curves. 
Fig.~\ref{fig:substitution_results}\subref{subfig:substitution_results_landscape} shows the results of this analysis. We observe that each 
motif has a characteristic genomic landscape, which helps in its biological interpretation.
For example, occurrences of Motif 13 are enriched in exons and conserved elements compared to hot regions in general -- their lowered substitution rates may indeed correlate with such enrichments.

%% file: discussion.tex
This article, for the first time to the best of our knowledge, tackles the problem of \emph{functional motif discovery} from a statistical perspective.
We proposed probKMA for discovering candidate motifs in a set of curves, incorporating ideas from functional data analysis, bioinformatics and fuzzy clustering. 
In addition, we proposed a generalized silhouette index to evaluate probKMA results, and 
implemented a post-processing stage for merging candidate motifs and searching motif occurrences along the curves. Although many alternative strategies can be employed in post-processing, each with pros and cons, results on simulated and 
real data suggest that our implementation is effective in a range of scenarios.

ProbKMA employs a flexible definition of curve similarity, which incorporates both levels and derivatives. In addition, similarity is defined locally, in a way that tolerates large gaps in the curves. 
This broadens the application scope of our methodology.
ProbKMA can also be applied to multivariate curves and, importantly, does not require the user to specify motif lengths or variability levels at the outset. These are \emph{learned from the data}, substantially improving performance with respect to approaches where lengths and/or radii are fixed. 

In our experience, motif discovery with probKMA can fail when motifs are too similar to one another or when they are too similar to background portions of the curves. This can happen by chance when motifs are very noisy.
Relatedly, simulations show that, when motifs are very noisy and/or dispersed in very long curves, 
our method
can identify motifs that were not intentionally introduced in the data -- but rather randomly created when generating background portions of the curves. In a way, these additional motifs may be considered as ``unintentional'' and yet true (as opposed to false) positives; they do recur in the curves in a way that is detectable by the algorithm.  Nevertheless, in our simulations they are noisier 
and have fewer occurrences. 
This observation underscores the need for further work addressing the \emph{statistical significance} of motifs found by probKMA. Notably, the flexible model based on B-splines that we introduced to generate simulation data may play an important role in this context, providing a way to estimate the likelihood of discovering motifs in background curves. 

Separately from its motif discovery purpose, probKMA can also be employed for probabilistic clustering of misaligned functional data based on local similarities. In this respect, it also represents a generalization of sparse clustering procedures recently proposed in functional data analysis \citep{fraiman2016,floriello2017}. In the limit, when the minimum motif length is close to the length of the curves under consideration, probKMA becomes a probabilistic version of $K$-mean with (global) alignment \citep{sangalli2010}.

Code implementing our methodology in R is available upon request. An R package is in preparation. 